# $Bi^{3+}$ Doped Nanocrystalline Ni-Co-Zn Spinel Ferrites: Tuning of Physical, Electrical, Dielectric and Magnetic Properties for Advanced Spintronics Applications

Md. Mahfuzur Rahman[a], Nazmul Hasan[b, e], Sumaiya Tabassum[c], M. Harun-Or-Rashid[d], Md. Harunur Rashid[b], Md. Arifuzzaman[b ‡]

[a] Department of Industrial and Production Engineering, Bangladesh University of Textiles, Dhaka-1208, Bangladesh

[b] Department of Mathematics and Physics, North South University, Dhaka-1229, Bangladesh

[c] Department of Apparel Engineering, Bangladesh University of Textiles, Dhaka-1208, Bangladesh

[d] Department of Physics, Bangladesh University of Textiles, Dhaka-1208, Bangladesh

[e] Department of Electrical and Computer Engineering, University of Rochester, NY 14620, USA

‡Corresponding Author: md.arifuzzaman01@northsouth.edu

## Abstract

This study reports the synthesis and characterization of nanocrystalline $Ni_{0.5}Co_{0.2}Zn_{0.3}Bi_xFe_{2-x}O_4$ (x = 0.0, 0.025, 0.050, 0.075, 0.100) ferrites synthesized via the sol-gel auto combustion method. The structural, morphological, electric, dielectric, and magnetic properties of $Bi^{3+}$-doped Ni-Co-Zn spinel ferrites annealed at 700 °C and further sintered at 850 °C, have been investigated towards analyzing the effect of $Bi^{3+}$ doping. X-ray diffraction (XRD) patterns and Fourier Transform Infrared Spectroscopy (FTIR) spectra have revealed the single-phase cubic-spinel structure of all inspected materials while retaining their high crystalline nature. Their average crystallite size and average grain size are found in the nanoscale range (48-74 nm) and (46-67 nm), respectively. The saturation magnetization ($M_s$) and experimental magnetic moment ($\eta_{exp}$) are found to decrease with increasing $Bi^{3+}$ content. The samples sintered at 850 °C display higher AC resistivity, attributing to the reduction of electrons hopping through grains in the samples. The low coercivity values (23.68 – 87.71 Oe) are observed, classifying the investigated materials as soft ferromagnetic. The increased magnetic anisotropy (K) through $Bi^{3+}$ doping indicates tunable stability in magnetic orientations, making them suitable for multifunctional applications.

***Index Terms***: Nano-spinel Ferrites; Ni-Co-Zn Mixed Ferrites; $Bi^{3+}$ Doping; Magnetization; AC Resistivity.

## Necessary equations

| Parameters | Relations | No. |
|---|---|---|
| **Rotational permeability ($\mu_{rp}$)** | $\mu_{rp} = 1 + \frac{8\pi M_S}{3H_a}$ | (1) |
| **Magnetic anisotropy field** | $H_a = \frac{H_c}{0.48}$ | (2) |
| **Net magnetic moment** | $\mu_B = \frac{M_w \times M_S}{5585}$ | (3) |
| **Lattice spacing** | 2dsinθ=nλ<br>order of diffraction, n=1 | (4) |
| **Dielectric** – Dielectric loss tangent (*tan* $\delta_E$): | $tan\delta_E = \frac{1}{\omega\varepsilon_o\varepsilon' \rho}$ | (5) |
| **Dielectric** – The real part ($\varepsilon'$): | $\varepsilon' = \frac{Ct}{\varepsilon_o A}$ | |
| **Dielectric** – Imaginary part ($\varepsilon''$): | $\varepsilon'' = \varepsilon' tan\delta_E$ | |
| **Lattice parameter** | $a_0 = d_{hkl}\sqrt{(h^2+k^2+l^2)}$ | (6) |
| **Crystallite size** | $D = \frac{0.9\lambda}{\beta_{hkl}cos\theta}$ | (7) |
| **Dislocation density** | $\delta = \frac{1}{D^2}$ | (8) |
| **Lattice and Micro strain** | $\varepsilon_{ls} = \frac{\beta}{4\,tan\theta}$; $\varepsilon_{ms} = \frac{\beta cos\theta}{4}$ | (9) |
| **Stacking fault** | $SF = \frac{2\pi^2}{45\sqrt{3\tan\theta}}$ | (10) |
| **Packing factor** | $P = \frac{D}{d}$ | (11) |
| **Crystal and Bulk Density** | $\rho_{th} = \frac{8M_w}{N_a a_0^3}$; $\rho_{ex} = \frac{M}{\pi r^2 l}$ | (12) |
| **Porosity** | $P\ (\%) = \frac{\rho_{th} - \rho_{ex}}{\rho_{th}} \times 100\ \%$<br>P (%) = $P_{inter} + P_{intra}$ | (13) |
| **Ionic Radii of A and B Sublattices** | $r_A = \sqrt{3}a_0(u_o - 0.25) - r_o$<br>$r_B = a_0(0.625 - u_o) - r_o$ | (14) |
| **Tolerance Factor** | $T = \frac{1}{\sqrt{3}}\left(\frac{r_A + r_o}{r_B + r_o}\right) + \frac{1}{\sqrt{2}}\left(\frac{r_o}{r_A + r_o}\right)$ | (15) |
| **Hopping lengths** | $L_{A-A} = \frac{a_o\sqrt{3}}{4}$; $L_{A-B} = \frac{a_o\sqrt{11}}{8}$; $L_{B-B} = \frac{a_o}{2\sqrt{2}}$ | (16) |
| **Polaron Radius** | $\gamma_p = \frac{1}{2}\sqrt[3]{\frac{\pi a_o^3}{576}}$ | (17) |
| **Average Grain Size** | $G_a = \frac{1.5L}{XN}$ | (18) |
| **Dielectric Loss Tangent** | $tan\delta_E = (1-P)tan\delta_o + C_m P^n$ | (19) |
| **M-H Loop Relation** | $H_C = \frac{0.96 \times K}{M_S}$ | (20) |
| **Complex Electric Modulus** | $M^* = \frac{1}{\varepsilon^*} = \frac{1}{\varepsilon' - i\varepsilon''} = \frac{\varepsilon'}{\varepsilon'^2 + \varepsilon''^2} - i\frac{\varepsilon''}{\varepsilon'^2 + \varepsilon''^2} = M'(real) + iM''(imaginary)$ | (21) |

## 1. Introduction

In today's technological sphere, nanocrystalline spinel ferrites have unequivocally emerged as a focal point in scientific research due to their unique properties. These nano-sized magnetic materials exhibit distinct states of matter, characterized by their high surface area-to-volume ratios and profound quantum effects [1]. In this study, we shed light on a novel dimension of the research landscape by investigating the effect of $Bi^{3+}$ doping on the physical, dielectric, electric, and magnetic properties of nanocrystalline Ni-Co-Zn spinel ferrites.

Spinel ferrites (SFs) are incredibly versatile magnetic materials widely employed in high-frequency domains, such as microwave and radio technologies, transformers, electrical devices, sensors, high-density information storage media, communication systems, and medical applications. [2-4]. Spinel ferrites can be represented as $AB_2O_4$, with cations distributed across tetrahedral-A and octahedral-B sites. The configuration of tetrahedral (A-O) and octahedral (B-O) structures in spinel ferrites plays a pivotal role in altering their magnetic, dielectric, and electrical properties. The resulting cationic distribution has several significant attributes, including robust magnetic saturation, minimal coercivity, enhanced electrical conductance, low electrical losses, inherent non-toxicity, magnetocrystalline anisotropy, and intrinsic magnetic and electrochemical characteristics.[5-7]. Spinel ferrites exhibit exceptional magnetic, electrical, optical, and catalytic properties from the intricate interplay of composition, valence states, and super-exchange interactions between lattice sites. When divalent metal ions (e.g. $Mg^{2+}$, $Ni^{2+}$, $Cu^{2+}$, $Li^{2+}$) occupy the A-site and trivalent metal ions such as $Co^{3+}$, $Fe^{3+}$, $Al^{3+}$, $Bi^{3+}$, $Sc^{3+}$ are distributed over the B-site, a semiconductor nature is achieved by forming an effective network between these inter-sites through dynamical interaction [8]. The multifaceted properties of SFs are decisively influenced by a spectrum of factors, including synthesis methods, precursor choices, type of dopant metals, chemical composition, and annealing or sintering temperatures [9-10].

The magnetic materials research landscape is dynamic, driven by the distinct dielectric, magnetic, and electrical properties of nickel (Ni)-, cobalt (Co)-, and zinc (Zn)-based spinel ferrites. Over the last few decades, Ni-based spinel ferrites have been extensively studied and have been widely utilized in various applications, including electromagnetic devices and biomedical technologies, owing to their biocompatibility nature [11-14]. Notably, Co-based spinel ferrites are currently of scientific interest due to their exceptional magnetic and catalytic properties. They have great potential for use in catalysis, electromagnetic interference shielding, sensors, and magnetic device storage [15-17]. Recently, there has been a growing focus on research into Zn-based spinel ferrites due to their enhanced antibacterial properties, high electrical conductivity, improved electromagnetic characteristics, and structural stability. These materials are now being utilized in various technological fields, including environmental sensors, supercapacitors, batteries, solar cells, and other electronic devices [18-19]. Moreover, Co-doping in Zn-based nanoparticles enhances ferromagnetism and low-frequency dielectric properties, suggesting potential for spintronic applications with a tunable bandgap [20]. The strong exchange interactions among the magnetic ions (Ni, Co, Zn in the A site) give rise to ferrimagnetism, resulting in minimal dielectric loss and a predominant spin-rotation resonance, causing magnetic loss for absorption performance. Investigating mixed Ni-Co-Zn nano-ferrites is essential to comprehensively analyze their structural, morphological, electrical, photo-catalytic, and magneto-dielectric properties. [21], [22-25]. Mixed spinel Ni-Co-Zn ferrites have high permeability, saturation magnetization, and low coercivity, making them suitable for diverse industrial applications [26-28]. However, the intricate selection of metallic ions over A- or B- sites is influenced by factors such as ion radius, free energy minimization, sintering processes, and temperature conditions, which opens up the doors for various metallic ions, including $Mg^{2+}$, $Ni^{2+}$, $Cu^{2+}$, $Ce^{2+}$, $Zn^{2+}$, $Al^{3+}$, $Sc^{3+}$, $Bi^{3+}$, as substituents over A- and B- lattice sites. Mention that rare-earth ion doping in nano-spinel ferrites has a substantial impact on their structural, elastic, and dielectric properties, opening up new opportunities for tailoring ferrites through substitutions to meet the requirements of specific optoelectronic applications in industries [29]. Therefore, the proper selection of those substituents can effectively improve the structural, physical, and magnetic properties of mixed spinel ferrite nanoparticles [2], [4], [10], [15], [30], [30-36]. Modifying ferrites through strain engineering, such as substituting $Dy^{3+}$ in the B-site of spinel ferrites, creates tensile strain. This results in stronger interactions within the spinel structure, enhancing magnetism, maintaining the single-phase spinel structure, and facilitating magnetoelectric coupling. Consequently, these ferrites become suitable for low-power devices [37-38]. High-entropy spinel ferrites deliver superior energy storage performance with enhanced charge transfer behaviour due to the combined high capacity of Fe-oxidation and metallic network formation during cycling [39].

Non-magnetic bismuth ($Bi^{3+}$) doping has expanded the potential applications of spinel ferrites due to enhanced electrical conductivity, reduced magnetic anisotropy, and multifunctional capabilities. [40-41]. The untapped potential of bismuth-based ferrite materials in terms of their synergy with eco-friendly synthesis methods remains a subject worthy of further exploration. Their inherent biocompatibility enhances their allure, particularly in biomedical applications.

As referred to in [42], the incorporation of Bi doping in the B-site of Ni-Co-Zn spinel ferrite unequivocally enhances the energy storage capacity through multi-electron electrochemical transport, thereby resulting in markedly elevated energy and power density in supercapacitor devices. Additionally, the substitution of $Bi^{3+}$ in the B-site of Ni-Co spinel ferrites consistently yields ferrimagnetic phases, irrespective of temperature, as confirmed by the Mössbauer study [43]. Furthermore, this substitution demonstrates a temperature-dependent transition from soft to hard magnetic behaviour. The addition of bismuth (Bi) to the B-site of cobalt ferrites has been shown to enhance the magnetic parameters ($M_s$, $H_c$, $M_r$) for magnetic recording and memory applications in spinel ferrites [44]. The increase in $Bi^{3+}$ content within Ni-Co spinel ferrites leads to a simultaneous elevation in the magnetic anisotropy constant and the optical bandgap energy from 0.71 eV to 2.12 eV. This observation underscores these materials' dual ferromagnetic and semiconducting nature, rendering them well-suited for high magneto-recording devices [45]. The incorporation of $Bi^{3+}$ into Mn-based spinel ferrites leads to a non-Debye type space-charge relaxation, resulting in a decrease in magnetic parameters. This phenomenon has been attributed to Maxwell-Wagner interfacial-type polarization, indicating their potential for applications in photonics [46].

In sol-gel-derived nanocrystalline bismuth-doped nickel ferrites, there is a clear ferroelectric polarization hysteresis loop of 10 $\mu C/cm^2$, which results in a magnetodielectric constant of 2.8% at room temperature. This observation highlights the potential of these materials as an alternative multiferroic material for nonvolatile memory devices [47-48]. Recent research on $Bi^{3+}$ doped multiferroic Ni-Co-Zn materials has revealed superparamagnetic behaviour with an improved magnetoelectric coupling constant and low coercivity. These findings indicate their potential suitability for applications in spintronic devices [49]. Inorganic nanomaterials such as spinel ferrites exhibit highly adaptable physical and interfacial properties, rendering them well-suited for sensitive biosensors and molecular imaging. Their adjustability facilitates the creation of biologically responsive magnetic resonance (MR) bimodal and near-infrared (NIR) fluorescence probes to develop "on-demand" drug delivery systems that can be activated by external signals [50]. The biocompatibility of spinel ferrite nanoparticles containing $Bi^{3+}$ ions presents promising applications in biomedicine. These nanoparticles interact well with biological systems and possess magneto-electric properties, making them suitable for innovative therapeutic approaches such as targeted drug delivery, hyperthermia for cancer treatment, and imaging. As a result, there are enhanced opportunities for advancing medical treatments [51-53]. Raju et al. [54] investigated the impact of Zn substitution on the structural and magnetic properties of Ni-Co ferrites. They found that the maximum saturation magnetic moment was observed for $Ni_{0.5}Co_{0.2}Zn_{0.3}$.

Considering the multifunctional benefits of Bi doping in spinel ferrites, modifying the physical, dielectric, and magnetic properties of Ni-Co-Zn mixed spinel ferrites through $Bi^{3+}$ doping could yield novel and unexplored results in the existing literature. Henceforth, we aim to investigate the potentials of $Bi^{3+}$ doped $Ni_{0.5}Co_{0.2}Zn_{0.3}Bi_xFe_{2-x}O_4$ synthesized through a cost-effective and standard sol-gel process [35].

## 2. Sample preparation and characterizations

A series of $Ni_{0.5}Co_{0.2}Zn_{0.3}Bi_xFe_{2-x}O_4$ (x = 0.0, 0.025, 0.050, 0.075, 0.100) ferrite nanoparticles were synthesized using the sol-gel technique. Various metal nitrates of $Ni(NO_3)_2.6H_2O$, $Cu(NO_3)_2.3H_2O$, $Zn(NO_3)_2.6H_2O$, $Bi(NO_3)_3.9H_2O$, and $Fe(NO_3)_3.9H_2O$ were amalgamated with a few drops of distilled water and stirred using a magnetic agitator to achieve a uniform solution. The pH of the solution reached 7.0 with the gradual addition of $NH_4OH$, and the mixture was then heated at 80 °C for 1 hour. The temperature continued to rise at a rate of 5 °C per hour until reaching 95 °C to transform the fluid into a dry gel. The dried gel self-ignited and burned for about two to three minutes at 95°C. Afterwards, the remaining ash underwent annealing at 700 °C for 5 hours to create a highly crystalline structure. The ferrites were intentionally annealed at 700 °C and sintered at 850 °C to enhance their physical and magnetoelectronic properties. This carefully planned process was based on solid evidence from previous studies, which showed that crystallinity was achieved at 700 °C, eddy current and magnetic loss increased at higher sintering temperatures, and peak dielectric values were observed at 850 °C in ferrite materials [55-57]. The ash powder was crushed with a mortar and pestle for several hours to ensure homogeneity. Disk-shaped samples were then created to analyze dielectric and electromagnetic properties. The nanocrystalline powder was compacted into 12 mm diameter, 2-3 mm thick disk-shaped tablets using a hydraulic press at 65 MPa for 2 minutes and sintered at 850 °C. This process facilitated subsequent physical, dielectric, electrical, and magnetic measurements.

Various robust methodologies were utilized to characterize the prepared ferrite samples comprehensively. Structural properties were analyzed using an X-ray diffractometer (XRD; PW3040, Cu-Kα, λ = 1.5418 Å), which provided insights into lattice parameters ($a$), crystal size ($D$), theoretical and bulk densities ($\rho$), displacement density ($\delta$), lattice strain ($\varepsilon_{ls}$), micro-strain ($\varepsilon_{ms}$), and stacking faults within the crystal structure of investigated ferrite nanoparticles. The morphology of the ferrite samples was meticulously analyzed using the JEOL-JSM 7600F model Field Emission Scanning Electron Microscope (FESEM). Energy-dispersive spectroscopy (EDS) confirmed the presence of Ni, Co, Bi, Fe, and O elements in each sample. Additionally, the electrical properties were assessed with the Wayne-Kerr impedance analyzer (6500B model), and the magnetic properties of the synthesized nanoparticles were investigated using a Vibrating Sample Magnetometer (VSM; Micro Sense, EV9). In characterizing the magnetic flux dynamics within the synthesized $Bi^{3+}$ doped Ni-Co-Zn spinel nano-ferrites, the rotational permeability ($\mu_{rp}$) has been assessed using equation (1) that considers the impact of saturation magnetization ($M_S$) and anisotropy field ($H_a$), both of which reflect the material's rotational response of magnetic moments under applied magnetic field [10]. The stability of magnets is crucial for several reasons, including controlled switching and demonstrating the efficacy of examined Ni-Co-Zn ferrite nanoparticles for real-world device uses. Consequently, the anisotropy field, calculated using equation (2), plays a crucial role, with magnetic coercivity ($H_c$) directly affecting the material's behaviour. The magnetic moment in spinel ferrites represents the natural magnetic strength resulting from aligned magnetic dipoles within the crystal. This is determined using equation (3), in which the material's molecular weight ($M_w$) and magnetic saturation ($M_s$) are direct contributing factors. The electrical properties of the synthesized nanoparticles have been investigated utilizing an impedance analyzer. The real part ($\varepsilon'$), imaginary part ($\varepsilon''$), and dielectric loss tangent ($tan\ \delta_E$) of dielectric have been calculated by relations contained in equation (4), which characterize several physical parameters of the materials such as capacitance ($C$), applied field frequency ($f$), free-space permittivity ($\varepsilon_o$), thickness ($t$), and surface area of contact surface of the prepared materials in tablet form.

## 3. Results and discussion

### 3.1 Structure analysis

The XRD patterns of prepared $Bi^{3+}$ substituted Ni-Co-Zn ferrites sintered at 850 °C, as depicted in **Fig. 1**, exhibit distinct peaks corresponding to diffractions from crystal planes of (111), (220), (311), (222), (400), (422), (511), and (440). The distinct shape of these peaks without secondary peaks suggests a uniform distribution with high crystallinity, supporting the single-phase cubic structures of the analyzed ferrite materials [31]. Lattice spacing determined by the inter-spacing distance of crystal planes ($d_{hkl}$) through equation (4), has been employed to estimate the average crystallite size using Debye-Scherer's equation considering the highest peak diffracted from the plane of (*311*). Nelson-Riley technique has been applied to deduce the lattice parameter ($a_0$) and unit cell volume ($V$) of the samples, which are listed in **Table 1**. The lattice parameter ($a_0$) and average crystallite size ($D$) have been calculated using equations (6) and (7), where $\lambda$ represents the wavelength of X-ray, $\beta_{hkl}$ is the full width at half maximum (FWHM) of highly intense peak (311), $\theta$ is the Bragg's angle, and $d_{hkl}$ is the inter-spacing distance of adjacent crystal planes.

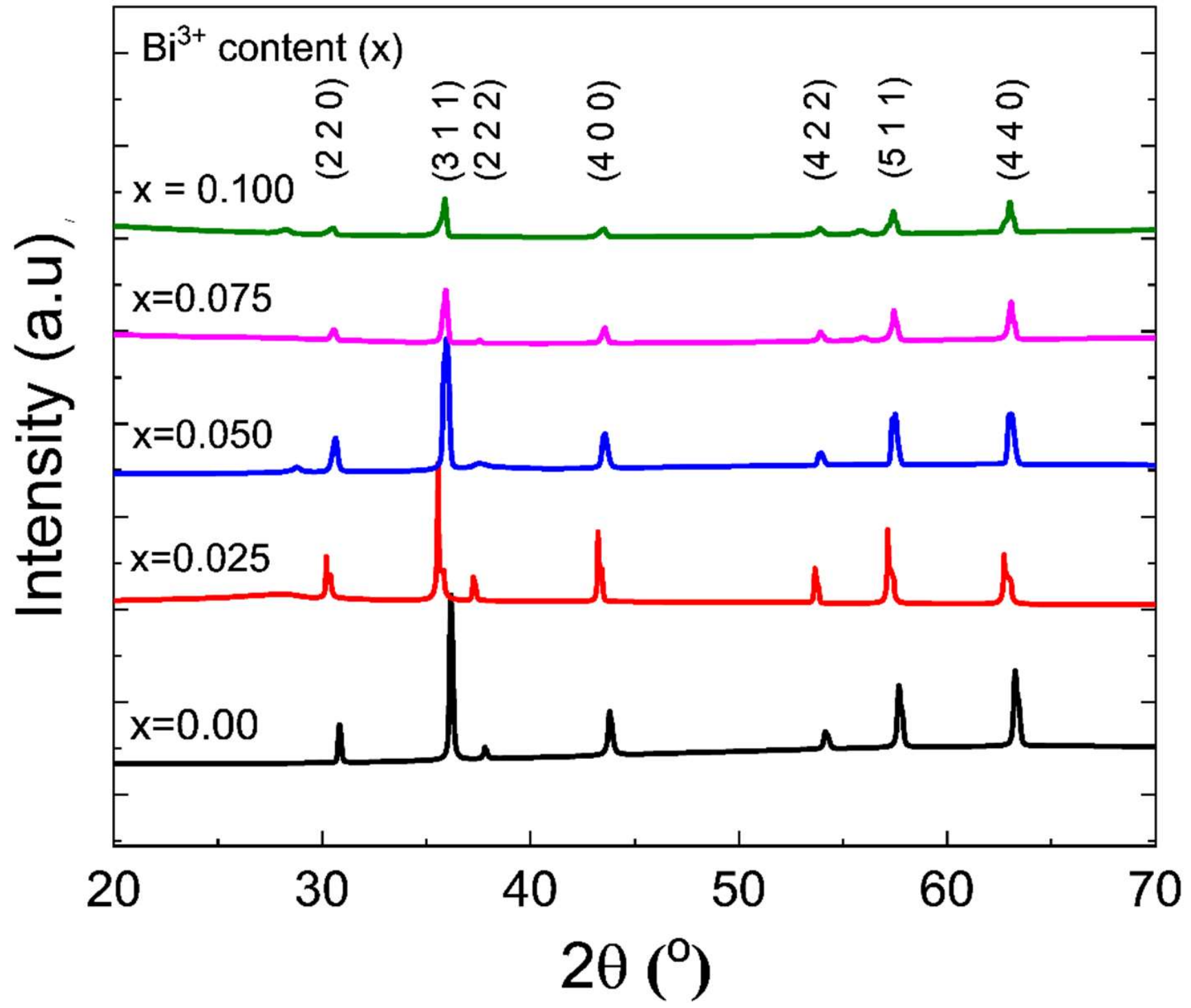

**Fig. 1.** XRD spectra of synthesized $Ni_{0.5}Co_{0.2}Zn_{0.3}Bi_xFe_{2-x}O_4$ sintered at 850 °C.

The lattice parameter increases with $Bi^{3+}$ doping concentration, ranging from 8.23 to 8.35 Å. Starting with $Bi^{3+}$ doping at x=0.025, both the lattice constant and cell volume exhibit a rising trend with $Bi^{3+}$ doping, attributing to the substitution of $Fe^{3+}$ (0.67 Å) cations with smaller radius by larger $Bi^{3+}$ (0.96 Å) ions. However, with the increased level of $Bi^{3+}$ doping, the lattice constant decreases due to the replacement of $Fe^{2+}$ ions (0.76 Å) by smaller radius $Bi^{5+}$ ions (0.74 Å). Besides, the cell volume ($V$) peaks at x=0.025 and decreases linearly with increasing $Bi^{3+}$ content. This trend is consistent with Vegard's law, indicating a significant reduction in the unit cell size [31]. As illustrated in Fig. 2(a), the average crystallite size demonstrates an initial rise with $Bi^{3+}$ content up to x=0.050, followed by a decrease with further $Bi^{3+}$ doping. This behaviour is attributed to the difference in ionic radius between $Bi^{3+}$ and $Fe^{3+}$, as well as between $Bi^{5+}$ and $Fe^{2+}$ ions, due to cationic redistribution over A and B-sites. This redistribution leads to an increase in lattice stress and strain [58].

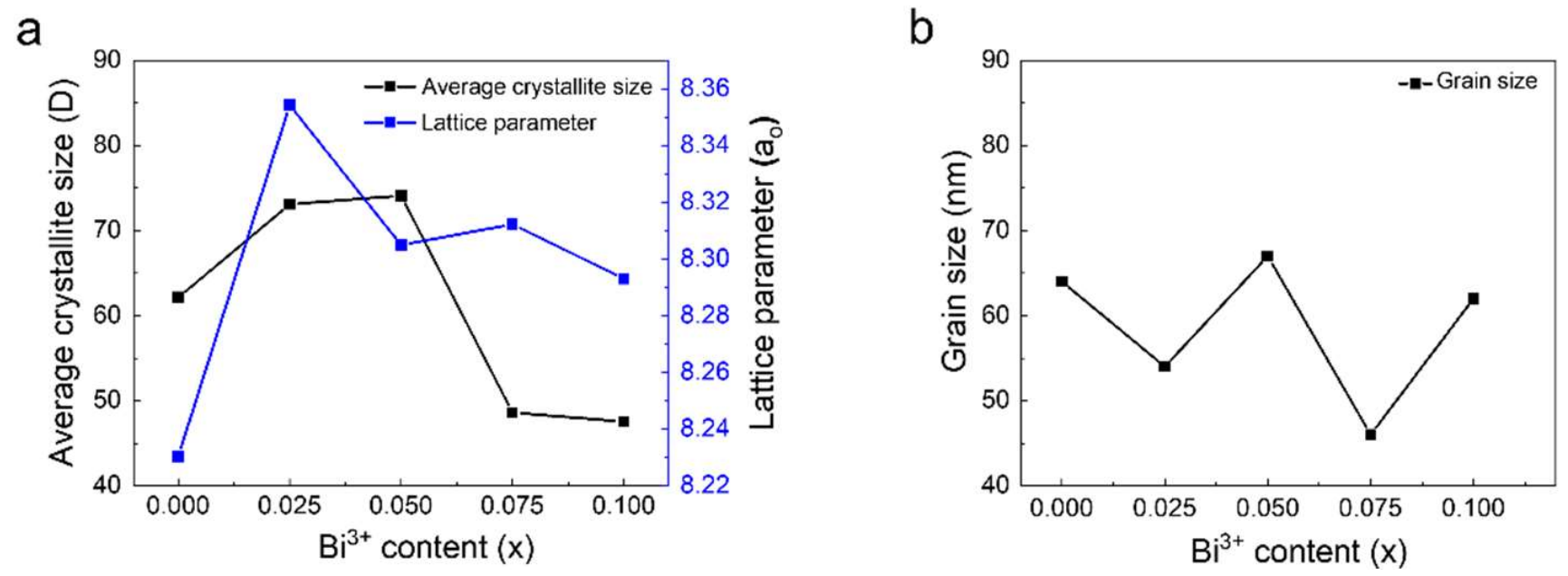

**Fig. 2.** Variation in (a) average crystallite size (*D*) and lattice parameter ($a_0$) and (b) grain size of the synthesized synthesized $Ni_{0.5}Co_{0.2}Zn_{0.3}Bi_xFe_{2-x}O_4$ varying with $Bi^{3+}$ content.

The dislocation density ($\delta$) is defined as the crystal structure's total displacement length per unit volume. It exhibits an inverse relationship with the particle size, characterized by a linearity error [3]. The dislocation density ($\delta$) of synthesized $Bi^{3+}$ doped Ni-Co-Zn spinel ferrites is determined using equation (8), and the calculated values can be found in **Table 1**. The reduction in $\delta$ corresponds to the pristine nature of the crystal lattice. In contrast, the increase of $\delta$ is attributed to additional defects or distortions. Lattice strain ($\epsilon_{ls}$) characterizes the distortion of atomic arrangement within spinel ferrite crystals, assessed through diffraction analysis given by equation (10), based on diffraction angle ($\theta$) and full width at half maximum ($\beta$). The lattice strain decreased from 9.72 (at x=0) to 8.58 (at x=0.050), indicating enhanced orderliness and reduced distortion due to increased $Bi^{3+}$ doping. This promotes a more regular atomic arrangement, improving the multiferroic properties of the synthesized nanoferrites [59]. However, the subsequent increase in lattice strain from 13.09 for x=0.075 to 13.36 for x=0.100 is observed due to an increase in lattice distortion due to lattice mismatch in the crystal. Spinel ferrites exhibit microstrain from crystal defects such as displacement and plastic deformation, occurring at approximately one part per million. This microstrain significantly impacts electrical conductivity, magnetic behaviour, and optical responses through phase transitions and lattice distortions in the crystal structure [60]. With increasing $Bi^{3+}$ content from x = 0 to x = 0.050, a decreasing trend in microstrain indicates a potential improvement in crystal orderliness and a reduction in lattice distortion. However, distortion or defects appear to rise at higher $Bi^{3+}$ concentrations (x = 0.075 and x = 0.100). This variation in microstrain provides insights into how the doping process influences the structural integrity of spinel ferrites with $Bi^{3+}$ doping.

The calculation of stacking faults in spinel ferrites is considered important as it is caused by lattice disorder and affects various properties such as charge carrier mobility, structural properties, electrical resistivity, phase transformations, ferromagnetic resonance (FMR) behaviour, and overall magnetic properties [61-62]. Thus, the stacking faults of $Bi^{3+}$ doped Ni-Co-Zn nano-spinel ferrites have been determined using equation (10) and listed in **Table 1**. Stacking faults (SF) have been found to decrease with $Bi^{3+}$ content from x= 0.025 to x = 0.050, thus improving the ferromagnetic resonance (FMR) properties through the reduction of local magnetic disorder [63]. Reducing stacking faults to a level as low as x = 0.050 may result in fewer spin-polarised electron scattering centres, improving spin coherence and relaxation time. Conversely, increased stacking faults can enhance spin transport dynamics at high doping concentrations (specifically, x = 0.075 and x = 0.100) due to additional scattering centres [56]. The packing factor (*P*) represents the degree of atomic arrangement in the crystal lattice of $Bi^{3+}$ doped Ni-Co-Zn mixed spinel ferrites [64], which has been estimated using equation (11). Its' higher values indicate a more ordered crystal lattice with elevated electrical conductivity [65]. Additionally, a well-organized lattice with a higher packing factor improves magnetic interactions in the studied ferrite nanoparticles, affecting the electron mobility of the systems [2]. Moreover, the packing factor contributes to structural stability; its higher values augment the stability and reduce the susceptibility [66]. The packing factor (*P*) variation with $Bi^{3+}$ content in nanocrystalline Ni-Co-Zn ferrites is shown in Table 1. The packing factor decreases at x=0.050 $Bi^{3+}$ doping, indicating lattice distortion or defect creation in the crystals. Conversely, a doping concentration above 5% results in a higher packing factor. The crystal density ($\rho_{th}$), bulk density ($\rho_{ex}$), and porosity (*P*) have been derived from molecular weight calculations as shown in **Table 1**. The difference between the theoretical and experimental densities reveals the impact of porosity on structural integrity, mechanical strength, thermal conductivity, and electrical behaviour [27]. Higher porosity within the crystal lattice of the investigated ferrites indicates the existence of open spaces or voids. These structural characteristics can significantly impact the thermal, mechanical, magnetic, and electrical properties as well as the overall durability of the material [67]. The introduction of $Bi^{3+}$ doping significantly enhances the material's porosity, resulting in a remarkable increase in specific surface area and reactivity, making it an excellent choice for catalysis or gas-sensing applications [8].

The interaction of ionic radii and the tolerance factor offers opportunities to customize spinel ferrites for specific applications by adjusting their structural and functional attributes. When ionic radii are mismatched, they create lattice strain, defects, and distortions, impacting electronic energy states and resulting in the quantum confinement effect in spinel ferrite crystals [68].

**Table. 1:** Structural parameters of synthesized nanocrystalline $Ni_{0.5}Co_{0.2}Zn_{0.3}Bi_xFe_{2-x}O_4$ ferrites.

| **Parameters** | **x=0.0** | **x=0.025** | **x=0.050** | **x=0.075** | **x=0.100** |
|---|---|---|---|---|---|
| d (Å) | 2.48154 | 2.51889 | 2.50401 | 2.50621 | 2.50042 |
| $a_0$ (Å) | 8.2303 | 8.3542 | 8.3049 | 8.3122 | 8.293 |
| V $(Å)^3$ | 557.503 | 583.06 | 572.80 | 574.31 | 570.34 |
| D (nm) | 62.11 | 73.08 | 74.10 | 48.60 | 47.53 |
| $\delta$ ($\times 10^{14}$) (lines/m$^2$) | 2.592 | 1.872 | 1.821 | 4.234 | 4.427 |
| $\varepsilon_{ms}$ ($\times 10^{-4}$) (line$^{-2}$/m$^{-4}$) | 5.28 | 4.747 | 4.682 | 7.138 | 7.30 |
| $\varepsilon_{LS}$ ($\times 10^{-2}$) | 9.72 | 8.75 | 8.58 | 13.09 | 13.36 |
| SF | 3.437 | 3.439 | 3.428 | 3.43 | 3.426 |
| P | 250.29 | 290.13 | 295.93 | 193.92 | 190.09 |
| $\rho_{ex}$ (kg/m$^3$) ($\times 10^3$) | 3.642 | 3.481 | 3.366 | 2.997 | 2.737 |
| $\rho_{th}$ (kg/m$^3$) ($\times 10^3$) | 5.6358 | 5.476 | 5.663 | 5.737 | 5.866 |
| P (%) | 35.38 | 36.43 | 40.56 | 47.76 | 53.34 |
| $G_a$ (nm) | 64.00 | 54.00 | 67.00 | 46.00 | 62.00 |
| $r_A$ (Å) | 0.4619 | 0.4887 | 0.4781 | 0.4796 | 0.4755 |
| $r_B$ (Å) | 0.7376 | 0.7686 | 0.7562 | 0.7581 | 0.7533 |
| $L_{A-A}$ (Å) | 3.564 | 3.617 | 3.596 | 3.599 | 3.591 |
| $L_{A-B}$ (Å) | 3.412 | 3.463 | 3.443 | 3.446 | 3.438 |
| $L_{B-B}$ (Å) | 2.910 | 2.954 | 2.936 | 2.939 | 2.932 |
| $\gamma_p$ (Å) | 0.7244 | 0.7353 | 0.7309 | 0.7316 | 0.7299 |
| T | 1.0238 | 1.01603 | 1.01911 | 1.01864 | 1.01984 |

**Table 1** lists the calculated values of ionic radii of A-sublattice ($r_A$), B-sublattice ($r_B$), and tolerance factor ($T$ ). When the T values are slightly above one, it indicates a slight distortion from the inverse spinel structure. As the concentration of $Bi^{3+}$ increases, the T values decrease, leading to a reduction in atomic distortion and an improvement in the stability of the inverse spinel structure.

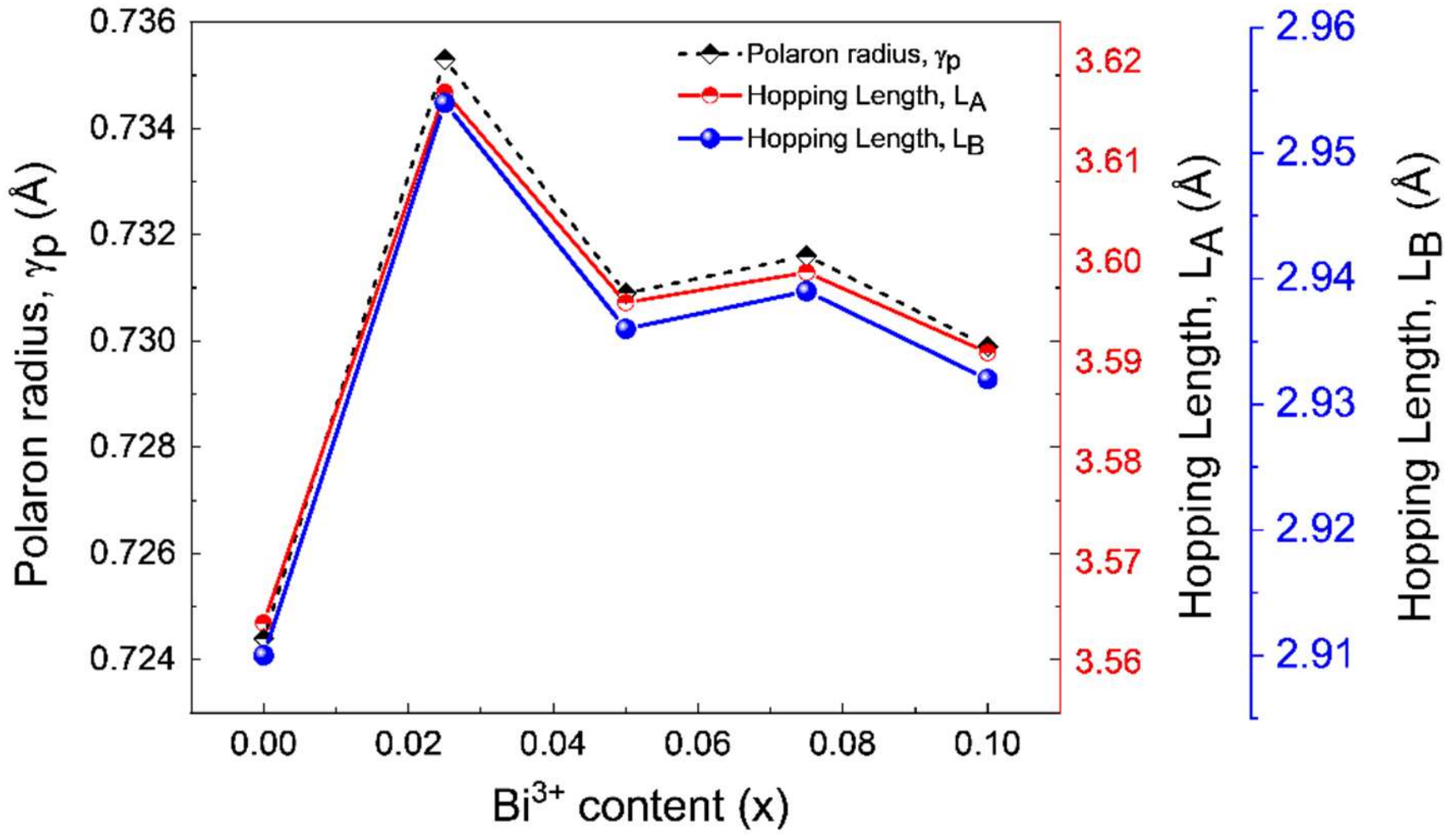

**Fig. 3.** Correlated polaron hopping behaviour of the synthesized $Ni_{0.5}Co_{0.2}Zn_{0.3}Bi_xFe_{2-x}O_4$.

The hopping length denotes the distance electrons cover between successive hopping events involving the exchange of valence states at octahedral sites. This process has the potential to alter the electrical conductivity properties of materials [69]. When polarons form, they cause localized electron-lattice distortions, which change the effective mass of charge carriers, ultimately impacting the performance of the material [70]. The dynamics of trapping and de-trapping, which are essential for evaluating electron transport mechanisms, are necessary for comprehending the behaviour of materials in photocatalytic activities [71]. Changes in hopping length also impact the distribution of ions within the crystal lattice, influencing the occupancy of tetrahedral and octahedral sites. The hopping lengths for A-A, B-B, and A-B sites are calculated using the following equations (16), where $a_0$ represents the lattice constant. Shorter hopping lengths facilitate efficient charge transport, enabling electrons to traverse shorter distances between tetrahedral and octahedral sites and promoting a more uniform distribution of charge carriers in the lattice, contributing to enhanced electrical conductivity [69]. As hopping lengths increase, more energy is required to transport charge carriers between cationic sites, while shorter hopping distances imply less energy for charge transport. Fig. 3 shows that increasing the $Bi^{3+}$ concentration ($x > 2.5\%$) decreases the hopping lengths. Changes in hopping rates and polaron effects may influence the material's magnetic behaviour. The interplay between electronic and magnetic properties can result in coupling effects. For instance, changes in conductivity may affect the magnetic ordering, providing avenues for tailoring multifunctional materials [72]. The polaron radius ($\gamma_p$) has been calculated using equation (17), which shows the rising trend with $Bi^{3+}$ doping up to $x=0.025$ and then demonstrates the decreasing trend with hopping lengths.

### 3.2 Surface morphology

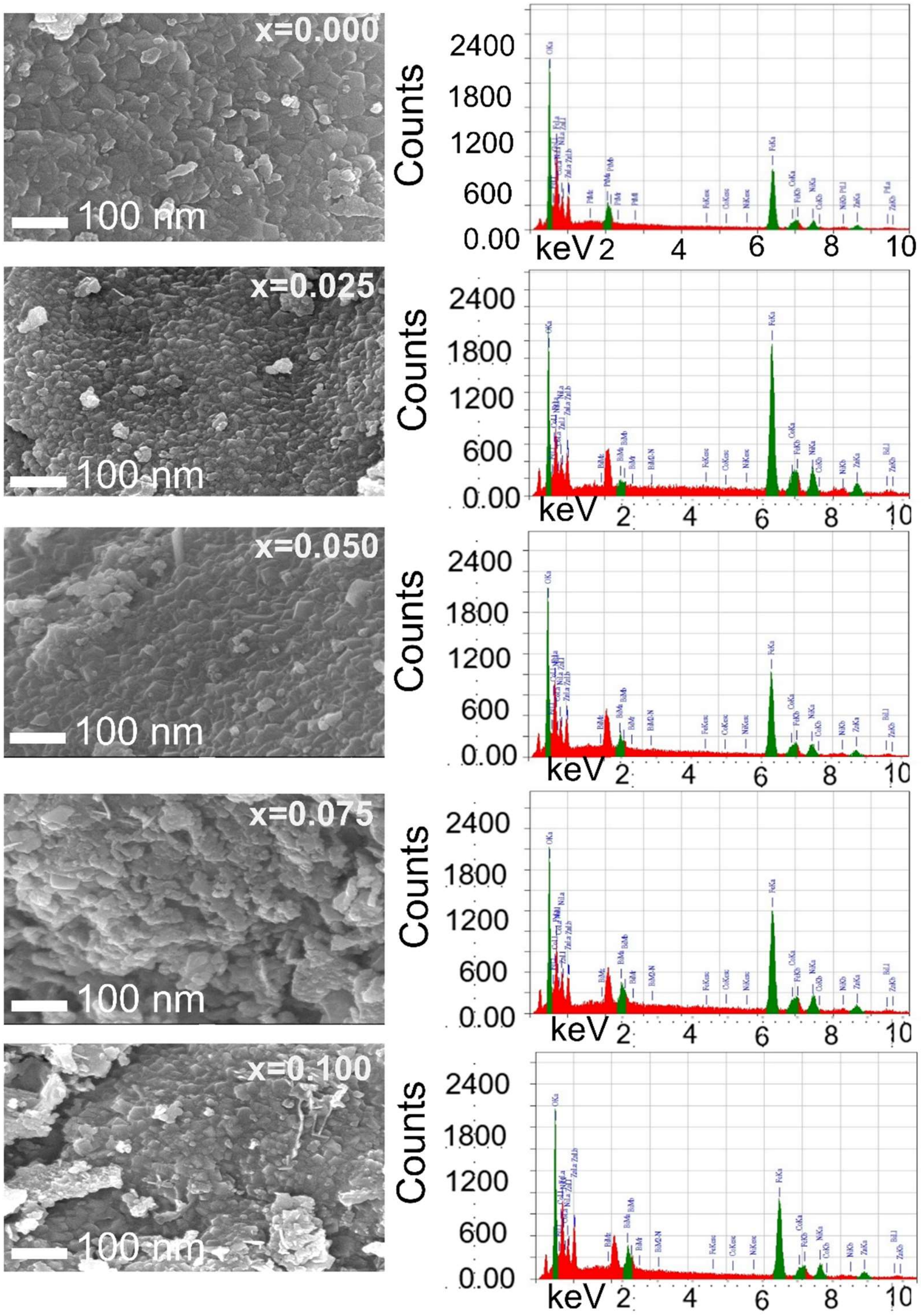

**Fig. 4.** Surface morphology micrographs using SEM (left column) and EDS analyzed compositional spectra (right column) of the synthesized $Ni_{0.5}Co_{0.2}Zn_{0.3}Bi_xFe_{2-x}O_4$.

**Fig. 4** shows the FESEM micrographs of $Ni_{0.5}Co_{0.2}Zn_{0.3}Bi_xFe_{2-x}O_4$ (x = 0.0, 0.025, 0.050, 0.075. 0.100) ferrites. The following relation computes the average grain ($G_a$) [3]:

$G_a = \frac{1.5L}{XN}$, where $L$ signifies the length of the test line in cm, $N$ is the number of intercepts, and $X$ represents the magnification of micrographs. The synthesized cubic spinel ferrites display relatively smooth textures with varying grain agglomeration degrees, as illustrated in Fig. 4. The observed grain sizes ranged from 46 to 67 nm. The micrographs show well-packed, crack-free granules with occasional cubic grain formation as x values increased. The FESEM micrograph indicates that all different $Bi^{3+}$ concentrations exhibited uniformity with well-defined grain boundaries. The brighter surface contrast indicates that $Bi^{3+}$ ions contributed to crystal rearrangement and changes in porous structures in the NiCoZn mixed spinel up to 10% doping concentration. [32]. The arrangement of grains and patterns can infer magnetic domain structures. The extracted FESEM images show that uniform grain alignment with consistent contrast variations suggests well-ordered magnetic moments in the synthesized Ni-Co-Zn ferrites. However, higher doping at 10% of $Bi^{3+}$ shows some stripe patterns that may lead to changes in magnetic domain configurations by interactions within the lattice [73].

Ferromagnetic resonance (FMR) analysis, correlated with particle size distribution, provides an understanding of magnetic properties influenced by quantum size effects [74]. Larger grains, as appeared in FESEM images with increasing $Bi^{3+}$ concentrations, may accommodate larger magnetic domains, contributing to strong ferromagnetic behaviour. However, lower $Bi^{3+}$ doping concentrations (up to 5%) exhibit homogenous smaller grains that may result in more localized magnetic behaviour that can impact the overall ferromagnetic resonance [63]. Moreover, major changes in contrast and the appearance of new features in the FESEM images for x=0.075 and 0.100 of $Bi^{3+}$ concentration may indicate phase transitions exhibiting distinct microstructural features [66]. The surface compositional elements of various $Ni_{0.5}Co_{0.2}Zn_{0.3}Bi_xFe_{2-x}O_4$ ferrites were investigated by the Energy-Dispersive X-ray Spectroscopy (*EDS*) analysis, as illustrated in **Fig. 4** (right panel for each correspondent material). The EDS spectra also confirm the existence of Ni, Co, Zn, Bi, Fe, and O elements in the synthesized spinel ferrites.

### 3.3 FTIR analysis

Spinel ferrites unequivocally showcase characteristic vibrational modes linked to metal-oxygen bonds' stretching and bending vibrations in the crystal lattice at precise frequencies. These modes are undeniably detected in the Fourier Transform Infrared Spectroscopy (FTIR) spectrum. **Fig. 5** illustrates the FTIR spectra for $Ni_{0.5}Co_{0.2}Zn_{0.3}Bi_xFe_{2-x}O_4$ samples sintered at 850 °C. Generally, in spinel ferrites, two prominent absorption bands, $\upsilon_1$ and $\upsilon_2$, correspond to the intrinsic stretching vibrations of M-O bonds at the tetrahedral (A-site) and octahedral (B-site) sites, respectively. An absorption peak at higher wavenumber υ1 (500-600 cm−1) is due to intrinsic stretching for A-site vibrations. Another lower-frequency weak band $\upsilon_2$ (350-400 $cm^{-1}$) appeared due to B-site vibrations in the measured spectra. **Fig. 5** also illustrates two distinct absorption peaks in the frequency range of 300-700 $cm^{-1}$ rising from the spinel sublattices, wherein the $\upsilon_1$ band experiences a shift from 533 to 538 $cm^{-1}$, indicative of enhanced stretching vibrations in tetrahedral metal-oxygen bonds. Simultaneously, the presence of lower wavenumber bands signifies weaker octahedral stretching effects. The observed frequency bands from both crystal sites in the FTIR spectra confirm the successful doping of $Bi^{3+}$ in nickel-cobalt-zinc ferrites and the formation of cubic spinel phases for all the samples according to Waldron's theory [73]. **Fig. 5** also reveals a slight $\upsilon_1$ band shift towards higher frequency, attributed to an increased $Bi^{3+}$ content in the Ni-Co-Zn ferrites. The stretching vibrations of $Fe^{3+}$ - $O_2$- bonds at both A- and B-sites definitively influence the lattice parameter. Initially, the lattice constant consistently demonstrates higher values, followed by a discernible decline with increasing $Bi^{3+}$ content. This pattern is confidently attributed to the impactful nature of the stretching vibrations of $Fe^{3+}$ - $O_2$- bonds on bond positions within the structure. The weaker absorptions are observed in the 1000-1250 $cm^{-1}$ range, and their intensity decreases as the concentration of $Bi^{3+}$ increases. This phenomenon is attributed to H-O-H bending vibrations, indicating changes in the chemical compositional density due to $Bi^{3+}$ doping concerning the crystal packing factor.

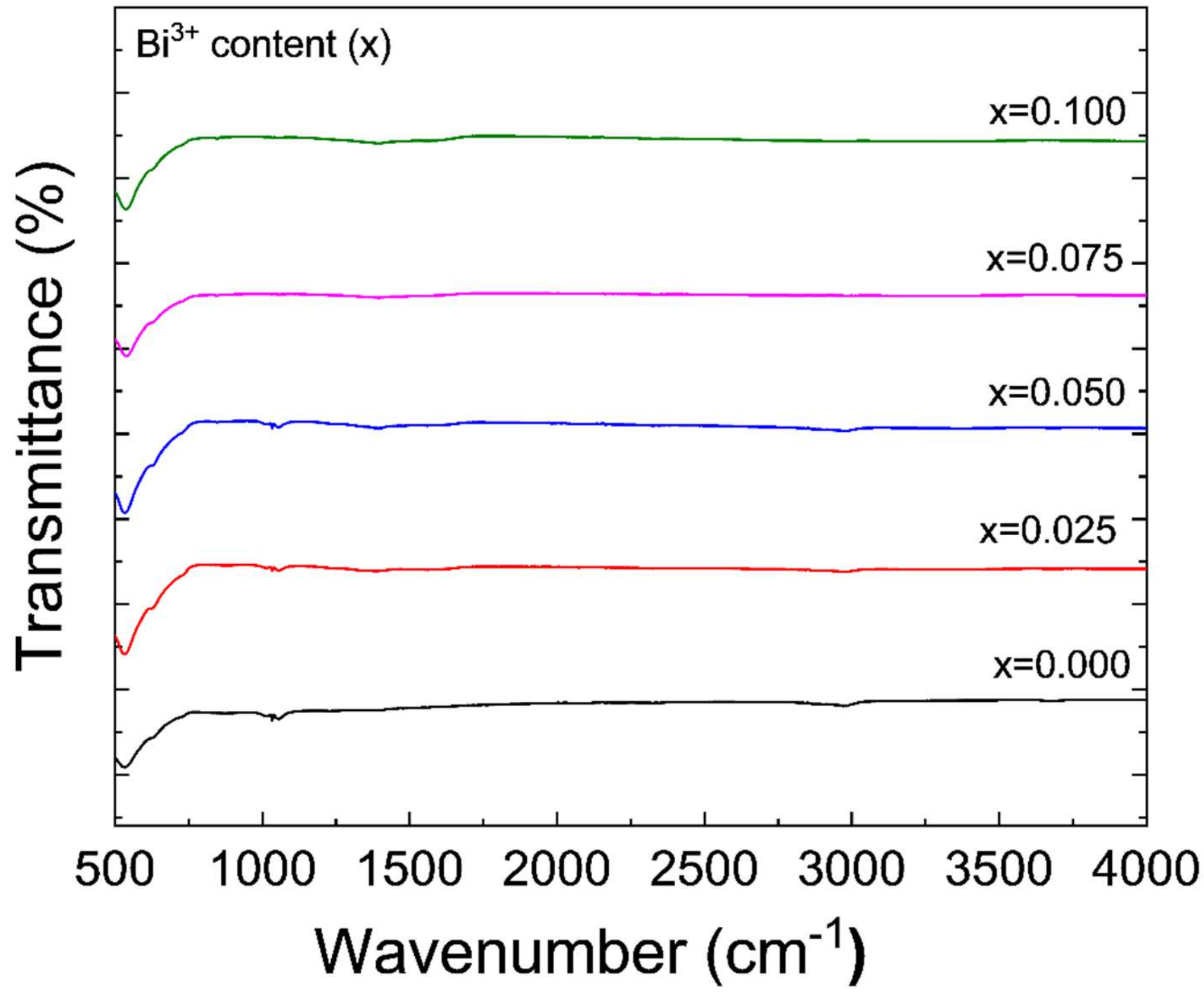

**Fig. 5.** FTIR absorption spectra of the synthesized $Ni_{0.5}Co_{0.2}Zn_{0.3}Bi_xFe_{2-x}O_4$ varying with $Bi^{3+}$ content.

### 3.4 Dielectric properties of the synthesized $Bi^{3+}$ doped $(NiCoZn)Fe_2O_4$ ferrites

#### *3.4.1 Dielectric constant and loss*

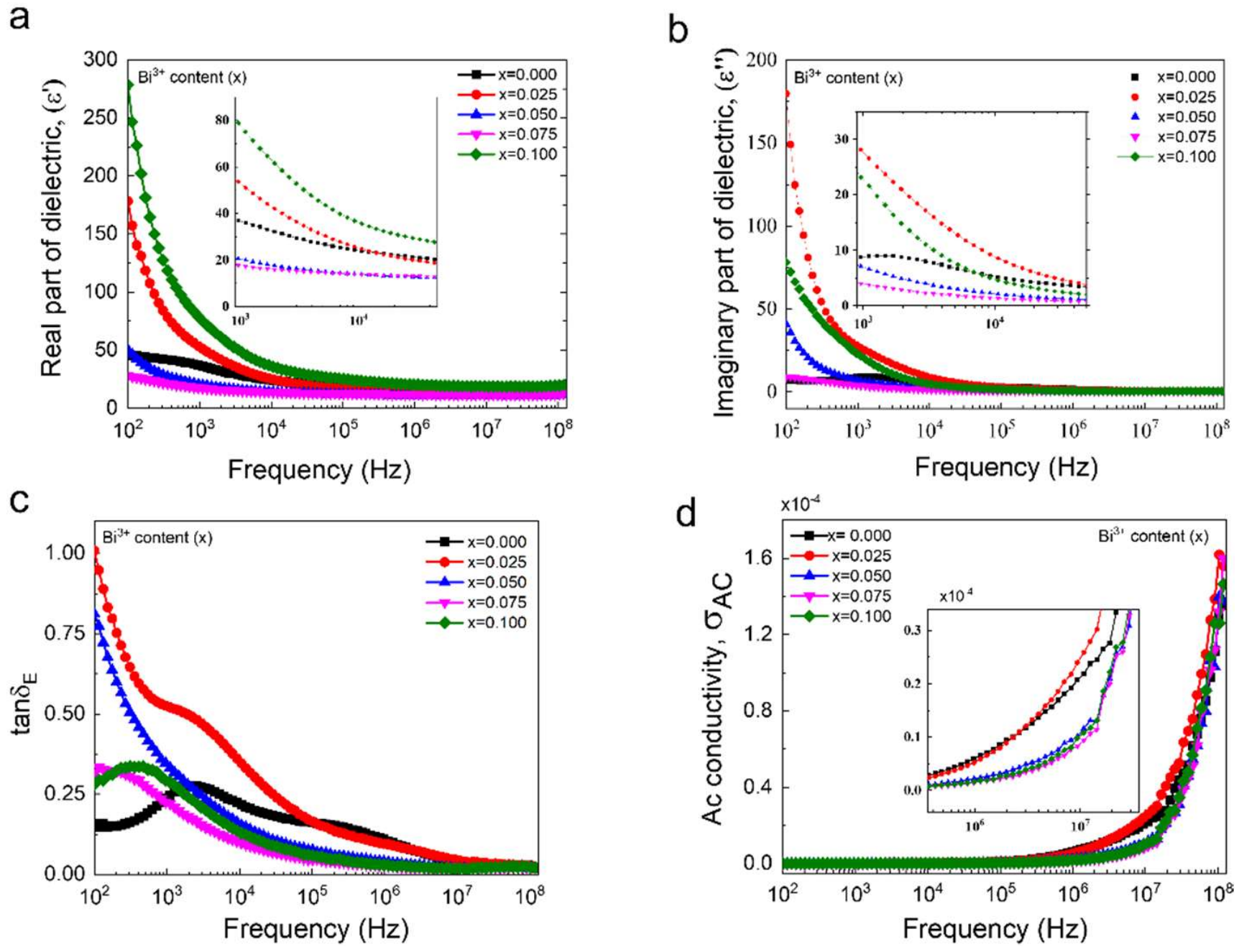

**Fig. 6.** Dielectric behaviour led to real (a) and imaginary (b) permittivity, dielectric loss tangent (c), and AC conductivity (d) of the synthesized $Ni_{0.5}Co_{0.2}Zn_{0.3}Bi_xFe_{2-x}O_4$.

Crystal structure, polarization mechanisms, and the magnetic-dielectric relationship influence the dielectric properties of spinel ferrites. Figure 6 illustrates the frequency-dependent complex dielectric response of the synthesized materials ($\varepsilon$), showing the variation in the real part of the dielectric constant. ($\varepsilon'$) part and imaginary ($\varepsilon''$) part for $Ni_{0.5}Co_{0.2}Zn_{0.3}Bi_xFe_{2-x}O_4$, sintered at 850 °C. The combination of these two components determines the overall dielectric behaviour of the materials, expressed by the relationship $\varepsilon = \varepsilon' - j\varepsilon''$, where the real part signifies the energy stored in a dielectric material, and the imaginary term characterizes the dissipated energy. In the lower frequency range, up to $10^4$ Hz, a more pronounced dielectric dispersion is evident in **Fig. 6 (a) & (b)**. In contrast, this behaviour becomes nearly independent of the applied field at higher frequencies. This dielectric dispersion in the ferrite is explicable through space charge polarization based on the Maxwell-Wagner two-layer model and Koop's phenomenological theory [73]. With an increasing concentration of $Bi^{3+}$, both the real and imaginary components of the dielectric constant exhibit higher values from 25 to 250 and 5 to 175, respectively, at $10^2$ Hz, which makes them advantageous as dielectric materials [75]. A notable decrease is observed up to $10^5$ Hz, followed by dielectric dispersion at higher frequencies, reflecting the ferromagnetic behaviour of the $Bi^{3+}$ doped Ni-Co-Zn spinel ferrites under investigation [76]. These dielectric properties render the studied materials suitable for various high-frequency applications, including energy absorbers and spintronics devices. As shown in **Fig. 6(a)**, the elevated resistivity of grain boundaries contributes to the higher values ε' over the low-frequency region. The increase in resistivity gives rise to space charge polarization, providing clarity on the electrical response in the synthesized spinel ferrites. At high frequencies, the electron hopping aligned with the electric field leads to a noticeable polarisation reduction, causing a decrease in the dielectric constant, as illustrated in Fig. 6(a). This reduction primarily stems from the decreased effectiveness of dipole polarization at higher frequencies. [77]. The value of (ε') shows the maximum for $Bi^{3+}$ content (x = 0.100) at lower frequencies. This can be attributed to $Bi^{3+}$ occupying A-sites in $Ni_{0.5}Co_{0.2}Zn_{0.3}Bi_xFe_{2-x}O_4$ and increased hopping between $Fe^{2+}$ and $Fe^{3+}$ charges at B-sites. The enhanced hopping activity triggers the accumulation of $Fe^{3+}$ ions within the grain and at the grain boundary, which is also evident in the morphological FESEM micrographs shown in Fig. 4. Consequently, there is a noticeable increase in space charge polarization, leading to a higher dielectric constant. In practical terms, the high flow of electricity in dielectric materials generates heat, contributing to material loss as characterized by the imaginary part of the dielectric constant ($\varepsilon''$). Heat dissipation is crucial for understanding the electrical behaviour of spinel ferrites. As depicted in **Fig. 6(b)**, there is a notable increase in the value of the dielectric loss ($\varepsilon''$) with increments of $Bi^{3+}$ content. This observed increase in ($\varepsilon''$) diminishes as the frequency rises, primarily influenced by the high resistivity at the grain boundaries. At higher frequencies, the frequent reversal of electron motion hinders the hopping of electrons to be aligned with the applied electric field. This dynamic leads to a reduced probability of charge accumulation at the grain boundary, resulting in reduced polarization [69]. Hence, the value of $\varepsilon''$ seems insignificant and stays consistent. Additionally, the porosity of the ferrite samples also affects their dielectric properties, with denser samples (as shown in Table 1) showing a higher dielectric constant value. Dielectric loss, a phenomenon resulting from impurities and imperfections in the crystal structure, manifests as a polarization lag behind the applied alternating field [34], [78]. **Fig. 6(c)** shows the variation in dielectric dissipation factor- loss tangent ($tan\delta_E$) of the samples sintered at 850 °C, varying frequencies. Fig. 6(c) depicts the increase of $tan\delta_E$ with $Bi^{3+}$ content at lower frequencies. The decreasing trend of dielectric loss tangent ($tan\delta_E$) with frequencies follows Koop's phenomenological theory. The material can efficiently convert electrical energy into thermal energy and generate heat, indicated by the dielectric loss tangent at higher frequencies. This phenomenon demonstrates a consistent pattern of normal dispersion due to Maxwell-Wagner polarization across all synthesized samples. The peak value of $tan\delta_E$ occurs at $\omega\tau = 1$, where $\omega$ is derived with the peak maximum frequency, $\tau$ stands for the relaxation time, which is inverse of twice the hopping probability (1/2P), and both parameters are closely linked to the hopping or jumping probability. Efficient electron sharing between $Fe^{3+}$ and $Fe^{2+}$ is achieved with minimal energy, reaching its maximum peak when the hopping frequency aligns with the applied electric field. The value of $tan\delta_E$ reaches its peak at lower conductive grain boundaries, demonstrating greater spontaneity than grains in the low-frequency region. Importantly, there is an energy loss during electron sharing between $Fe^{3+}$ and $Fe^{2+}$, necessitating a high energy input. [10].

### *3.4.2. AC Conductivity*

The frequency-dependent AC conductivity ($\sigma_{ac}$) in spinel ferrites reflects the material's ability to conduct electric current in response to an alternating electric field, making them useful for various device applications, including electromagnetic, microwave, and magnetic storage. **Fig. 6(d)** illustrates the variation in AC conductivity ($\sigma_{ac}$) with frequencies for the synthesized samples sintered at 850 °C. The electrical characterization of these nano ferrites, elucidated by the hopping mechanism, is closely tied to ($\sigma_{ac}$) followed by the relation, $\sigma_{ac} = \omega\varepsilon'\varepsilon_0 tan\delta$, where $\omega$ is the angular frequency. According to the hopping mechanism, electrons tend to migrate within the same elements of the sample, favouring dispersion across the B sites of the lattice. As in **Fig. 6(d),** $\sigma_{ac}$ of the investigated ferrites attains a greater magnitude at higher frequencies and diminishes as the frequency decreases. Subsequently, beyond a certain frequency, $\sigma_{ac}$ shows frequency-independency with a minimal magnitude, explicable by the frequency dependence of the grain and grain boundaries. At lower frequencies, the high-resistance grain boundary undergoes increased activation, impeding the free flow of charge. Consequently, the hopping of electrons between $Fe^{2+}$ and $Fe^{3+}$ ions occurs less frequently, resulting in a constant $\sigma_{ac}$ value. At higher frequencies, the increased hopping of electrons between $Fe^{2+}$ and $Fe^{3+}$ ions plays a critical role in elevating the $\sigma_{ac}$ value as charge carriers in spinel ferrites are getting localized and influenced by hopping mobility. The lower $\sigma_{ac}$ value at lower frequencies remains nearly constant for all compositions, primarily because the hopping frequency almost ceases to follow the applied field frequency. The rising $Bi^{3+}$ concentration in Ni–Co-Zn ferrites contributes to the linear increment in AC conductivity, involving enhanced valence ion exchanges between $Bi^{3+}/Bi^{5+}$ and $Fe^{2+}/Fe^{3+}$. The observed linear rise in AC conductivity with frequency in the synthesized $Bi^{3+}$ doped NiCoZn spinel ferrites substantiates the presence of polaronic conduction. In the small polaron model, conductivity experiences a linear augmentation with increasing frequency, whereas in the case of large polarons, conductivity diminishes with rising frequency. Conversely, at high frequencies, conduction takes place through low-resistance grains. This analysis establishes a correlation between AC conductivity, ion exchange, and the microstructural features of Ni–Co-Zn ferrites.

### *3.4.3. Complex electric modulus*

Introducing $Bi^{3+}$ through doping significantly perturbs the spinel ferrite lattice, infusing additional charge carriers and inducing intricate alterations in its electrical characteristics, rooted in the dynamic aspects of electrical transport phenomena. The intricate physics of the complex electric modulus in spinel ferrites offers an all-encompassing comprehension of their dynamic electrical behaviour. It provides indispensable insights into charge carrier dynamics, dielectric relaxation processes, and the influence of external factors such as doping and temperature. The complex electric modulus ($M^*$) is a frequency-dependent parameter that amalgamates the capacitive ($M'$) and resistive ($M''$) components of the material's response. In spinel ferrites' crystal lattice, dielectric relaxation occurs due to the movement of charge carriers, specifically the hopping of electrons between different valence states of metal ions. The relationship between dielectric properties and real and imaginary parts of electric modulus can be assessed for synthesized samples using equation (21).

**Fig. 7** represents room temperature dispersions of the complex electrical modulus behaviour for the synthesized spinel ferrites. The capacitive real part of electric modulus M′ of ferrites responds very well to higher frequencies for the $Bi^{3+}$ doped NiCoZn mixed spinels, as observed from the modulus spectrum in **Fig. 7(a)**. A discernible magnitude shift is evident among the doped samples, except for x=0.100, particularly at $10^4$ Hz. Notably, the highest value of M′ is observed for x=0.075 up to the frequency of $10^4$ Hz, and beyond this frequency, the sample with x=0.050 exhibits the highest modulus. The frequency-dependent M′ exhibits a monotonous asymptotic dispersion behaviour up to $10^4$ Hz, suggesting charge carriers' short-range mobility during conduction [33]. **Fig. 7(b)** illustrates the variation in the resistive component of the complex electric modulus (M″) with frequency and varying with $Bi^{3+}$ content. The peaking behaviour, elucidated by the hopping mechanism, provides a more accurate depiction of the transition of charge carriers in the materials. Fig. 7(b) shows that charge carriers are involved in the hopping conduction process over the long-range at low frequencies. Conversely, charge carriers hope at higher frequencies over short distances, indicating a relaxation in the polarization process. The M″ exhibits strong and symmetric peaks, and their positions shift, varying with frequency and $Bi^{3+}$ concentrations.

The Cole-Cole plots of the M′ vs. M″ spectrum were examined to resolve the contribution of grains and grain boundaries in the relaxation process of the synthesized $Ni_{0.5}Co_{0.2}Zn_{0.3}Bi_xFe_{2-x}O_4$ ($0 \leq x \leq 0.100$) at room temperature, as illustrated in **Fig. 7(c)**.

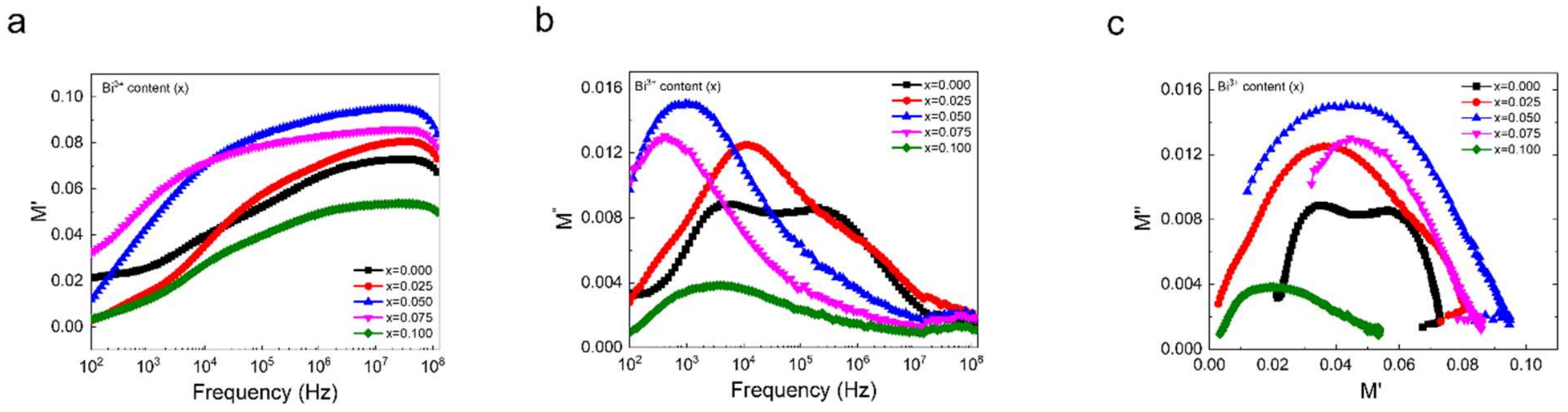

**Fig. 7.** Complex electric modulus plots (a) capacitive M′ dispersion with frequency, (b) resistive M″ frequency dependent behaviour, and (c) Cole−Cole plot M″ vs M′ Cole−Cole plot of electric modulus for the synthesized $Bi^{3+}$ doped $Ni_{0.5}Co_{0.2}Zn_{0.3}Bi_xFe_{2-x}O_4$ ($0 \leq x \leq 0.100$).

Usually, two semicircles represent multiple relaxation mechanisms within the material due to the capacitance of grains and grain boundaries capacitance, contributing to the conduction mechanism in spinel ferrites. Fig. 7(c) shows that the modulus spectrum has semicircular arcs with two lobes for the undoped Ni-Co-Zn spinel ferrites. Additionally, the presence of a single semicircle coupled with bulk properties for each $Bi^{3+}$ concentration in the synthesized NiCoZn mixed spinel ferrites indicates the impact of doping on relaxation time. With $Bi^{3+}$ content of x=0.100, a single asymmetric semicircular arc is observed, indicating the system's non-Debye type relaxation response. The shift of semicircles from the origin is noticed with frequencies and varying with different $Bi^{3+}$ content. Besides, $Bi^{3+}$ doping broadens the semicircles in the Cole-Cole plot, as shown in **Fig. 7(c).** This indicates a higher distribution of relaxation times, causing multiple relaxation mechanisms in the materials. This broadening suggests a more heterogeneous or complex dielectric response, reflecting the material's diverse charge carrier dynamics with possible lattice interactions. The observed electric modulus behaviour suggests a temperature-dependent hopping mechanism for charge transport in the studied spinel ferrite compounds, coupled with the non-Debye type dielectric relaxation [58].

### 3.5 Complex impedance analysis

The impedance analysis aims to discern the contribution of conductivity originating from both the resistive and reactive components under the influence of an applied AC field. Tuning impedance properties in spinel ferrites optimizes the performance of electronic devices and sensors. This modulation allows customization of temperature sensitivity, making them ideal for specific temperature-dependent applications, and provides control over the conduction mechanism. Furthermore, the Nyquist plot offers a thorough insight into the contributions of microstructural-induced resistivity and interfacial resistance at conducting electrodes. The frequency-dependent real ($Z'$) and imaginary ($Z''$) components of the complex impedance modulus, along with the Nyquist plot ($Z''$ vs. $Z'$), are illustrated in **Fig. 8 (a-c)**. The real component of the complex impedance ($Z' = R_s$) for the synthesized ferrites with varying $Bi^{3+}$ concentrations sintered at 850 °C is presented in **Fig. 8a** as a function of frequency. This plot of Z' with varying frequencies demonstrates the peak values at lower frequencies, followed by a monotonous decline with the increase of frequencies up to $10^4$ Hz. Beyond that frequency, the curves converge and become almost frequency-independent. This supports the observed frequency-dependent trend in AC conductivity above in **Fig. 6**, confirming the semiconducting characteristics of these ferrites [8], [34]. These lower-frequency impedance properties indicate their potential as dielectric materials in low-frequency devices. Besides, $Bi^{3+}$ doping increases impedance loss up to x=0.075, followed by a decrease of x=0.100 compared to the non-doped $Bi^{3+}$ phase. Beyond the frequency of 100 kHz, impedance diminishes for all prepared samples, signifying the dominance of the conduction mechanism in the ferrites attributed to the electron hopping at higher frequencies.

The frequency-dependent imaginary part of complex impedance ($Z'' = 1/\omega C_p$) for ferrites sintered at 850 °C, correlating with the pellet capacitance, is illustrated in **Fig. 8(b)**. The value of Z '' reaches the peak value (Z '' max) for $Bi^{3+}$ substitution at x=0.075. The observed peak signifies the dominance of grain boundary resistivity compared to the grains' contribution. After a frequency of 104 Hz, both Z' and Z'' converged at zero, which suggests the potential dissipation of interfacial charges for homogeneous phases in the samples. Overall, the decreased values of Z' and Z'' at high frequencies indicate the temperature-dependent multiple relaxation behaviour in the materials, devoid of an ideal singular relaxation time.

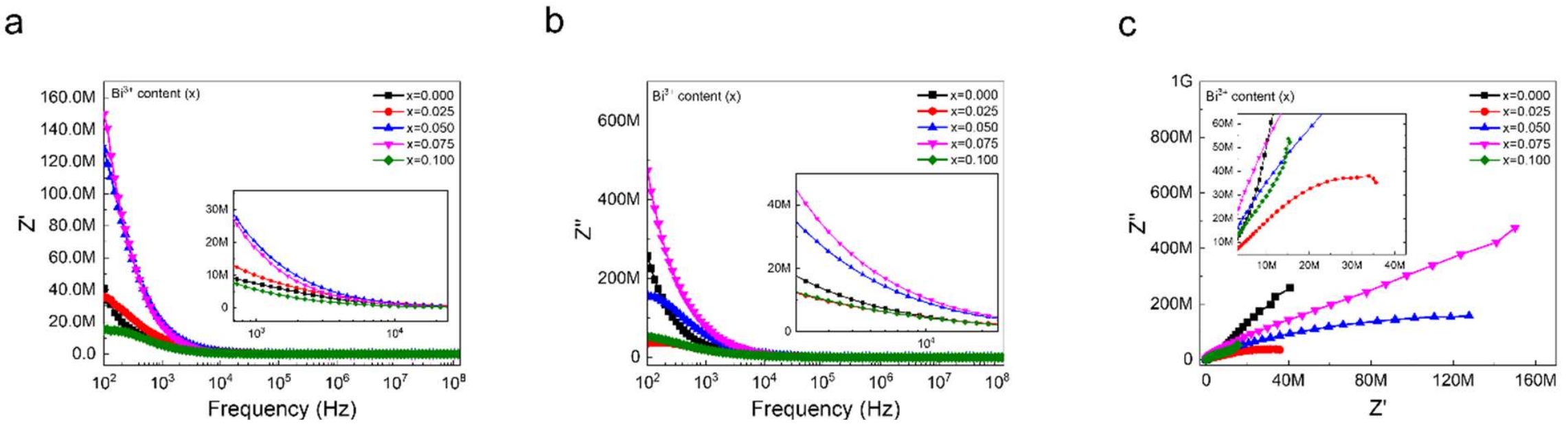

**Fig. 8.** Frequency-dependent (a) real part, (b) imaginary part, and (c) Cole−Cole plot Z″ vs Z′ for the complex impedance modulus of the synthesized $Ni_{0.5}Co_{0.2}Zn_{0.3}Bi_xFe_{2-x}O_4$ ($0 \leq x \leq 0.100$).

The Nyquist impedance plot in **Fig. 8(c)** effectively resolves the contribution of grain, grain boundary, and interfacial resistance of conducting electrodes in devices. A semicircular arc is prominently visible in the Nyquist impedance plot of Fig. 8(c), positioned below the real axis's centre, indicating the single-phase nanocrystalline behaviour. The size of this semicircular arc is found to decrease with $Bi^{3+}$ content, attributed to the grain boundary resistance. The distinctive nature of complex impedance spectra and the presence of various electrical responses are well-explained by the partial semicircular arcs, similar to the previous study [79]. The reduction of the radius of semicircular arcs in the Nyquist plot with increasing $Bi^{3+}$ concentration indicates a shift in the DC conduction due to grain boundaries. This suggests the enrichment of electrical conductivity with enhanced charge transport properties, which aligns with the electrical mechanism characterized by a single relaxation feature [80]. The undoped Ni-Co-Zn spinel ferrite phase exhibits the largest semi-circle at high frequency, indicating the highest $R_{ct}$ value. In contrast, the shorter semi-circles are apparent at higher frequencies with high $Bi^{3+}$ content for lower values of $R_{ct}$, indicating the improved electron transfer kinetics at the electrode/electrolyte interface. The materials exhibit a more pronounced Warburg line at low frequency at x=0.050 and 0.075, indicating the enhanced pseudocapacitive behaviour, resemblance an ideal capacitor with low $Z_w$, implying the capacitive charge storage kinetics [13], [42]. Increasing $Bi^{3+}$ content improves the mobility of space charge, enabling the accumulation of charge carriers near the phase boundaries to overcome barriers, enhancing the conductivity and reducing the impedance [48]. The decrease of semicircles arc indicates good electrode contact materials with increasing $Bi^{3+}$ content in the Ni-Co-Zn spinel ferrites. This type of behaviour aligns with the room-temperature multiferroic materials that tend to have phase transitions suitable for energy-storage applications [80]. Also, tunability in this impedance behaviour with electrode/electrolyte, diffuse later resistance, and device internal resistance for devices using Ni-Co-Zn spinel ferrites through $Bi^{3+}$ doping leads to the possibility of using the materials for Quantum Dot LED devices and High Performances Supercapacitors [81], [82].

The modulation of dielectric polarization impacts complex impedance, affecting the electrical responses in materials, particularly in the context of electronics and sensors. In spintronics, the manipulation of electron spin plays a crucial role in the efficiency of devices. The polarization in spinel ferrites, especially concerning grain and grain boundary resistances, indicates potential dielectric and relaxation phenomena. Dielectric relaxation processes, associated with the movement of charge carriers, can influence electron spin dynamics. [48]. Variations in complex impedance, especially the imaginary part (Z"), at different frequencies provide insights into polarization mechanisms, including the interplay between charge carriers and crystal lattice structure affecting spin behaviour. To harness this for electron spin manipulation, careful material design is needed. Understanding relaxation times associated with polarization can guide spintronic device development. External factors like temperature, doping, and microstructure should be considered for tailored spin manipulation. An interdisciplinary approach merging spintronics with dielectric and impedance studies holds promise for innovative electron spin control and information processing applications.

### 3.6 Magnetic properties

Magnetization in spinel ferrites is intricately governed by the arrangement of magnetic ions, predominantly transition metal ions, within the spinel crystal lattice. These magnetic ions, usually situated at the B-site, possess spin moments arising from unpaired electrons. Superexchange interactions, modulated by oxygen ions and governed by the face-centred cubic crystal structure and arrangement of metal ions, tortuously influence spin alignment, determining the material's magnetic characteristics. The VSM method is utilized at room temperature to evaluate field-dependent magnetic parameters, including saturation magnetization ($M_s$), coercivity ($H_c$), remanent magnetization ($M_r$), magnetic anisotropy constant ($K_a$), and experimental magnetic moment ($\eta_{exp}(\mu_B)$), as presented in **Table 2**, based on the M-H hysteresis loop of $Bi^{3+}$ doped Ni-Co-Zn ferrite nanoparticles depicted in **Fig. 9**.

The hysteresis loops illustrated in **Fig. 9(a)** depict the soft ferromagnetic behaviour and unequivocally demonstrate magnetization saturation at high applied field values of 11,000 Oe. Regardless of the composition of the nanoparticles, the M-H curves at room temperature display

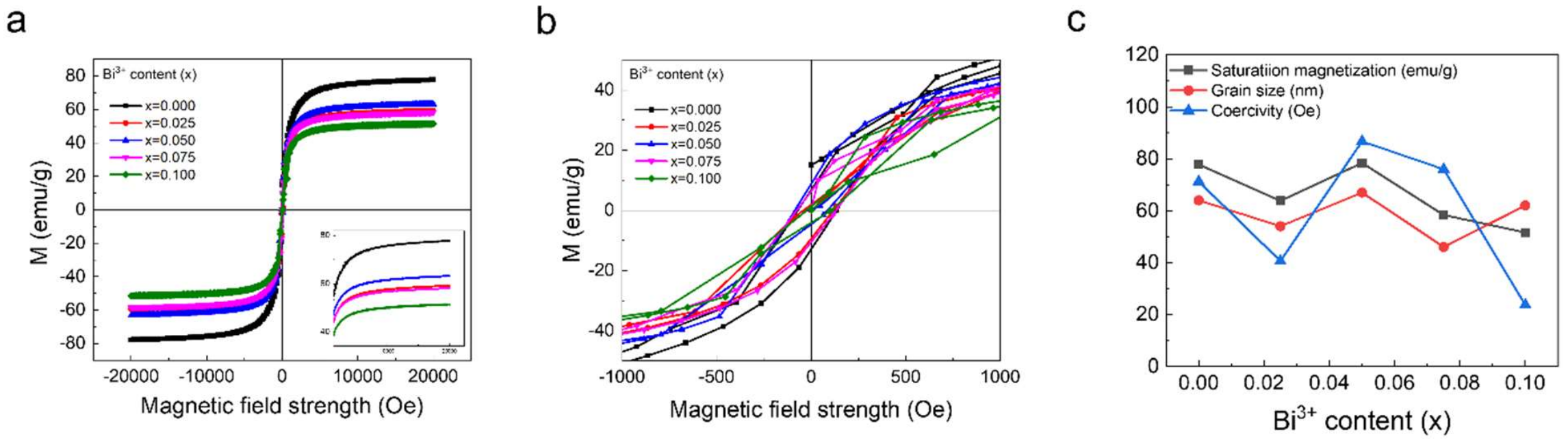

**Fig. 9.** Magnetic hysteresis loop (M-H) (a, b), and (c) variations in $M_s$, $H_c$ with G for the synthesized $Ni_{0.5}Co_{0.2}Zn_{0.3}Bi_xFe_{2-x}O_4$ ($0 \leq x \leq 0.100$).

S-shaped loops with remanence ($M_r$) and coercive field ($H_c$) corroborate the ferromagnetic behaviour. The asymmetry in cation distribution over A- and B-sites influences the variation in magnetic characteristics. The inset of **Fig. 9(a)** reveals that $Bi^{3+}$ doping leads to an initial increase in coercivity at x=0.025, followed by a subsequent decrease. This variation may be correlated with lattice strain and differences in grain structure, as indicated by the structural analysis and **Fig. 9(b)**. The rise in $Bi^{3+}$ concentrations resulted in closed M-H loops, signifying a decrease in coercivity ($H_c$) and remanent magnetization ($M_r$) within the ranges of 6.76 to 1.18 emu/g and 71.22 to 23.68 Oe, respectively. This suggests the potential tuning of diverse NSFs toward superparamagnetic transition at higher $Bi^{3+}$ concentrations [17]. The increased saturation magnetization ($M_S$) values with $Bi^{3+}$ concentration variations, particularly at x=0.0 and x=0.050, signify robust magnetic moments, enabling controlled magnetization in spintronic devices essential for applications such as magnetic memory. $Bi^{3+}$ doping-dependent values of the $H_c$, $M_r$, and squareness ratios (SQR = $M_r/M_s$) are presented in **Table 2**. Generally, SQR values ≥0.5 signify the presence of a single magnetic domain (SMD) structure, while SQRs less than 0.5 indicate the existence of multiple magnetic domains (MMDs).

The saturation magnetization decreases almost linearly with increasing $Bi^{3+}$ concentration, ranging from 77.78 emu/g at x=0 to 51.58 emu/g at x=0.100, except for x=0.050, where it reaches the highest value of 78.38 emu/g. This phenomenon can be elucidated based on the occupancy of magnetic and non-magnetic ions in tetrahedral and octahedral sites. The magnetic moments of various cations are as follows: $Ni^{2+}$ (2 $\mu_B$), $Co^{2+}$ (3.88 $\mu_B$), $Zn^{2+}$ (0 $\mu_B$), $Bi^{3+}$ (0 $\mu_B$), and $Fe^{3+}$ (5 $\mu_B$). The substitution of $Bi^{3+}$ for $Fe^{3+}$ at the octahedral B site leads to a decrease in saturation magnetization, impacting the moment at the tetrahedral site and inducing a B-B interaction that weakens magnetization at the octahedral site. $Bi^{3+}$ at the B site introduces spin canting and antiparallel spin coupling, diminishing the A-B exchange contact, consistent with the Yafet-Kittel model [83]. Consequently, the magnetic moment ($M_{cal}$) has been determined using the sub-two-lattice model proposed by Neel, expressed as $M_{cal} = M_{octa}(B) - M_{tetra}(A)$. Additionally, the saturation magnetization values exhibit a more flattened trend with increasing $Bi^{3+}$ compared to the undoped sample. The flatter $M_s$ curve with $Bi^{3+}$ content indicates enhanced stability against thermal influences, and the decrease in $M_s$ values suggests a more ordered magnetic state with reduced susceptibility to thermal disruptions [45]. Moreover, the magnetic saturation of the synthesized ferrites is influenced by particle size, as evident in **Fig. 9(b)**. Smaller grains with larger surface-to-volume ratios exhibit increased spin disorder, particularly when the surface-to-volume ratio is significant.

Experimental Magnetic Moment ($\eta_{exp}(\mu_B)$) provides insights into the collective behaviour of spins in addition to magnetization behaviour. The $\mu_B$ and $M_s$ values for the considered NiCoZn spinel ferrites peak at x = 0.050, signifying enhanced super-exchange interactions with the substitution of $Bi^{3+}$ ions. As the $Bi^{3+}$ x values increase, both $M_s$ and $\mu_B$ values decline, indicating a weakening of super-exchange interactions for higher substitution contents. Elevated $\eta_{exp}(\mu_B)$ at x=0.050 indicates tunable improved magnetic response suitable for effective spin manipulation and data storage in spintronics. All samples demonstrate ferromagnetic M-H loops with a moderate coercive field ranging from 23 to 86 Oe, indicative of the soft magnetic nature of these ferrites, facilitating easy magnetization reversal with low energy. The lower coercivity ($H_C$) values for x=0.025 and x=0.100 suggest relatively low resistance to changes in magnetic orientation, offering advantages for efficient switching in spintronic devices and reducing energy consumption. The magnetic anisotropy constant ($K$) describes the energy required to reorient the magnetic moments from their preferred axis, which signifies magnetic stability in materials. Higher values in magnetic anisotropy constant ($K$) in spinel ferrites indicate enhanced stability in magnetic orientation often favoured for stable long-term magnetization in spintronics and lower K values for applications requiring fast magnetic switchings like sensors or actuators. The K values in **Table 2** indicate favourable ranges across all samples, with a notable increase at x=0.050. This rise in K implies enhanced stability in magnetic orientations, crucial for maintaining coherent spin states, particularly in spintronic memory devices [47]. However, tailored values in SQR, μB, and K with $Bi^{3+}$ concentration variations also indicate their suitability as microwave absorbers, biomedical devices, and effective contrast agents for improved imaging.

| **Table. 2:** Magnetic parameters of the synthesized nanocrystalline **$Ni_{0.5}Co_{0.2}Zn_{0.3}Bi_xFe_{2-x}O_4$**. | | | | | | | | |
|---|---|---|---|---|---|---|---|---|
| **$Bi^{3+}$ content** | $M_S$ (emu/g) | $M_r$ (emu/g) | $H_C$ (Oe) | SQR=$M_r/M_S$ | K (erg/Oe) | $H_a$ | $\eta_{exp}(\mu_B)$ | $\mu_{rp}$ |
| **x=0.0** | 77.78 | 6.76 | 71.22 | 0.0869 | 5770.304 | 148.375 | 3.2927 | 9663 |
| **x=0.025** | 63.78 | 1.87 | 40.6 | 0.0293 | 2697.363 | 84.583 | 3.3461 | 4517 |
| **x=0.050** | 78.38 | 8.71 | 86.71 | 0.1111 | 7079.510 | 180.646 | 3.4257 | 11855 |
| **x=0.075** | 58.35 | 6.30 | 75.93 | 0.1080 | 4615.120 | 158.188 | 2.5902 | 7729 |
| **x=0.100** | 51.58 | 1.18 | 23.68 | 0.0229 | 1272.307 | 49.333 | 2.325 | 2131 |

Besides, the high $M_S$ signifies robust magnetic moments suitable for encoding quantum information, while tunable $H_C$ values are crucial for controllable qubit switching in quantum gates. Magnetic anisotropy (K) influences the stability of quantum states, contributing to sustained quantum entanglement. The tunability of magnetic properties through $Bi^{3+}$ doping provides an avenue for tailoring quantum bits, but practical implementation in QIP requires meticulous engineering to address these challenges and leverage the unique characteristics of the material for quantum computing applications [84], [85]. The combined electrical transport properties and magnetization dynamics of the $Bi^{3+}$ doped NiCoZn spinel ferrites demonstrate promising potential for advanced physical functionalities, such as high Tc superconductivity, magnetoresistance, topological quantum Hall effect, quantum spin liquid, etc., employing thin film engineering techniques [74], [86], [87], [88].

## Conclusion

$Bi^{3+}$-doped nanocrystalline Ni-Co-Zn mixed ferrites were successfully synthesized using the sol-gel auto-combustion method to investigate their physical, structural, morphological, magnetic, and electrical properties. The studied materials show a single-phase cubic spinel lattice structure, achieving a nanostructure with grain sizes ranging from 46-67 nm. A linear trend is observed in the lattice constants and crystallite size variation with increasing $Bi^{3+}$ doping. The grain and grain boundaries were analyzed through SEM analysis. The dielectric response of the materials resulting from $Bi^{3+}$ doping is due to the redistribution of cations over the sub-lattices. This redistribution facilitates efficient space charge accumulation, making the materials suitable for low dissipative electrical transport in practical devices. Additionally, the electrical and impedance properties mediated by $Bi^{3+}$ doping make these materials promising for multifunctional devices due to their multiferroic magnetoelectric coupling. The illustration of magnetization with varying $Bi^{3+}$ content demonstrates the ferromagnetic properties of nanocrystalline $Ni_{0.5}Co_{0.2}Zn_{0.3}Bi_xFe_{2-x}O_4$. The $Ni_{0.5}Co_{0.2}Zn_{0.3}Bi_{0.05}Fe_{1.95}O_4$ compound demonstrates exceptional magnetic properties, with the highest values of saturation magnetization (78.38 *emu/g*), coercivity (86.71 *Oe*), magnetic anisotropy constant (7079.51 *erg/Oe*), and experimental magnetic moment (3.43 $\mu_B$) observed for 5 % doping of $Bi^{3+}$. This positions the investigated materials as highly promising for advanced applications in spintronics. Comprehensive research and development efforts are imperative to fully exploit the remarkable potential of $Bi^{3+}$ doped Ni-Co-Zn spinel ferrites, paving the way for transformative advancements in spin-optoelectronic, biomedical, and quantum information technologies.

**CRediT authorship contribution statement**

**M. M. Rahman and N. Hasan**: Conceptualization, Methodology, Formal analysis, Investigation, Data curation, Software, Writing – original draft, review & editing, Visualization. **S. Tabassum**: Data Acquisition, Visualization, Writing –editing, Formal analysis, **M. Harun-Or-Rashid**: Writing – review & editing, Validation, Discussion and Formal analysis. **H. Rashid**: Review and editing, Validation, Project administration. **M. Arifuzzaman**: Conceptualization, Supervision, Formal analysis, Project administration, Critical revision, finalizing, submitting and correspondence of the manuscript. All the authors discussed, went through, and commented on the manuscript before submission.

**Declaration of Competing Interest**

The authors affirm that no known financial conflicts of interest could have influenced the work reported in this paper.

**Acknowledgments**

The authors thank the Center of Excellence of the Bangladesh Council of Scientific and Industrial Research and the Department of Mathematics and Physics at North South University (NSU), Dhaka 1229, Bangladesh. The NSU research grant, CTRG22-SEPS-10, partially supports this research.